\documentclass{eas}
\usepackage{graphicx}
\usepackage{natbib}
\usepackage{amsmath}
\usepackage{amssymb}

\begin{document}

\title{Mass loss of red supergiants: a key ingredient for the final evolution of massive stars}
\runningtitle{Mass loss of red supergiants}
\author{Cyril Georgy}\address{Astrophysics group, EPSAM, Keele University, Lennard-Jones Labs, Keele, ST5 5BG, UK}
\author{Sylvia Ekstr\"om}\address{Geneva Observatory, University of Geneva, Chemin des Maillettes 51, 1290 Versoix, CH}
%
%
\begin{abstract}
Mass-loss rates during the red supergiant phase are very poorly constrained from an observational or theoretical point of view. However, they can be very high, and make a massive star lose a lot of mass during this phase, influencing considerably the final evolution of the star: will it end as a red supergiant? Will it evolve bluewards by removing its hydrogen-rich envelope? In this paper, we briefly summarise the effects of this mass loss and of the related uncertainties, particularly on the population of blue supergiant stars.
\end{abstract}
\maketitle
\section{Introduction}

Mass loss is an important driver of the evolution of massive stars. Stars more massive than $\sim 15\,M_\odot$ lose during their life, $\sim4-7\,M_\odot$ (for a $15\,M_\odot$) to $\sim 100\,M_\odot$ \citep[for a $120\,M_\odot$ model, see][]{Ekstrom2012a,Chieffi2013a}. This loss of matter will progressively uncover deep layers of the star, modifying the surface chemical composition together with internal mixing processes. Stellar evolution models are currently not able to compute the mass-loss rate consistently, and thus rely on recipes, both observational or theoretical \citep[e.g.][]{deJager1988a,Kudritzki2000a,Vink2000a}

\begin{figure}
\begin{center}
\includegraphics[height=.27\textheight]{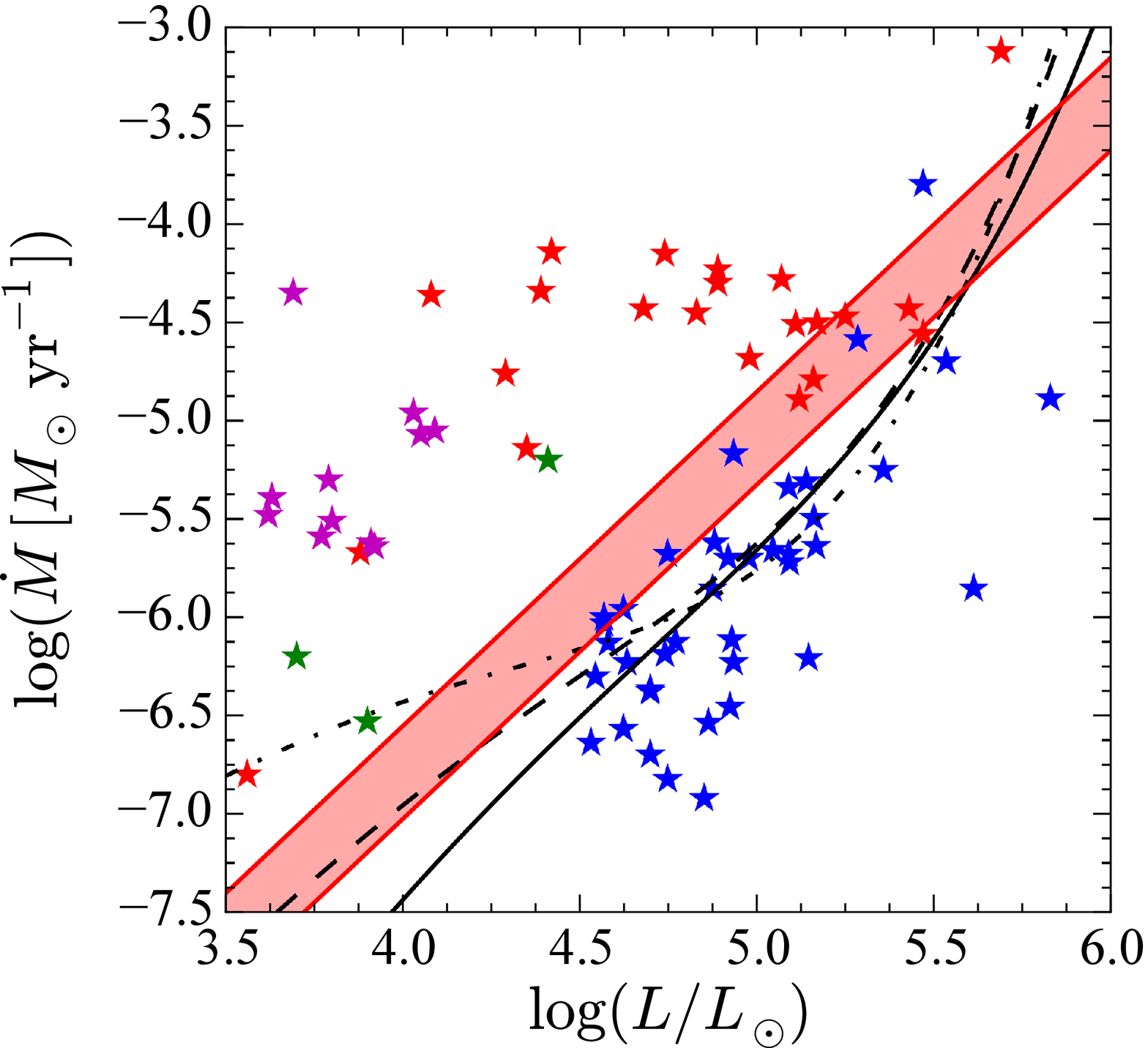}
\caption{Mass-loss rate during the RSG phase. Observations from \citet{Mauron2011a} (blue stars) and \citet{vanLoon2005a}: M-type stars (red stars), MS- and S-type stars (green), and carbon RSGs (purple). The black curves show the mass-loss rate according to \citet{deJager1988a}, for $\log(T_\text{eff}) = 3.5$ (solid), $3.6$ (dashed) and $3.7$ (dotted-dashed). The red zone correspond to the mass-loss rate used in the Geneva stellar evolution code for standard mass-loss rate (lower line), or increased one (top line).}
\label{Georgy_FigMdotRSG}
\end{center}
\end{figure}

One of the worse constrained of these rates is the red supergiant (RSG) one. As shown in Fig.~\ref{Georgy_FigMdotRSG}, the mass-loss rates have a huge scatter (three orders of magnitude).

In the latest release of stellar models grids published by the Geneva stellar evolution group \citep{Ekstrom2012a,Georgy2013b}, stars more massive than about $20\,M_\odot$ have layers in their envelope that have luminosity considerably higher than the Eddington one during the RSG phase. This could ease the superficial layers to escape the star. We have thus decided to increase the standard RSG mass-loss rates by a factor of $3$ in our computations for stars more massive than $20\,M_\odot$. This is shown by the red diagonal band in Fig.~\ref{Georgy_FigMdotRSG}, where the classical mass-loss rate of \citet{deJager1988a} is also plotted for comparison for three different effective temperatures. Our mass-loss rates are well in the range of the observed ones. In the following, we will discuss a few consequences of such an increased mass-loss rates during the RSG phase \citep[see also][]{Vanbeveren1998a}.

\section{Second crossing of the HRD}

As a direct consequence of the high mass-loss rates during the RSG phase, the stellar models lose much more easily their external hydrogen-rich envelope. This favours a bluewards motion in the HRD \citep{Giannone1967a}. In our models, this occurs for star more massive than $25\,M_\odot$ (non-rotating models) or $20\,M_\odot$ \citep[rotating ones, see][]{Georgy2012b}. Interestingly, it allows to fit the observed maximal luminosity of Galactic RSGs \citep{Ekstrom2012a}, as well as the maximal mass of observed type IIP supernova progenitors \citep{Smartt2009a}.

\begin{figure}
\begin{center}
\includegraphics[height=.2\textheight]{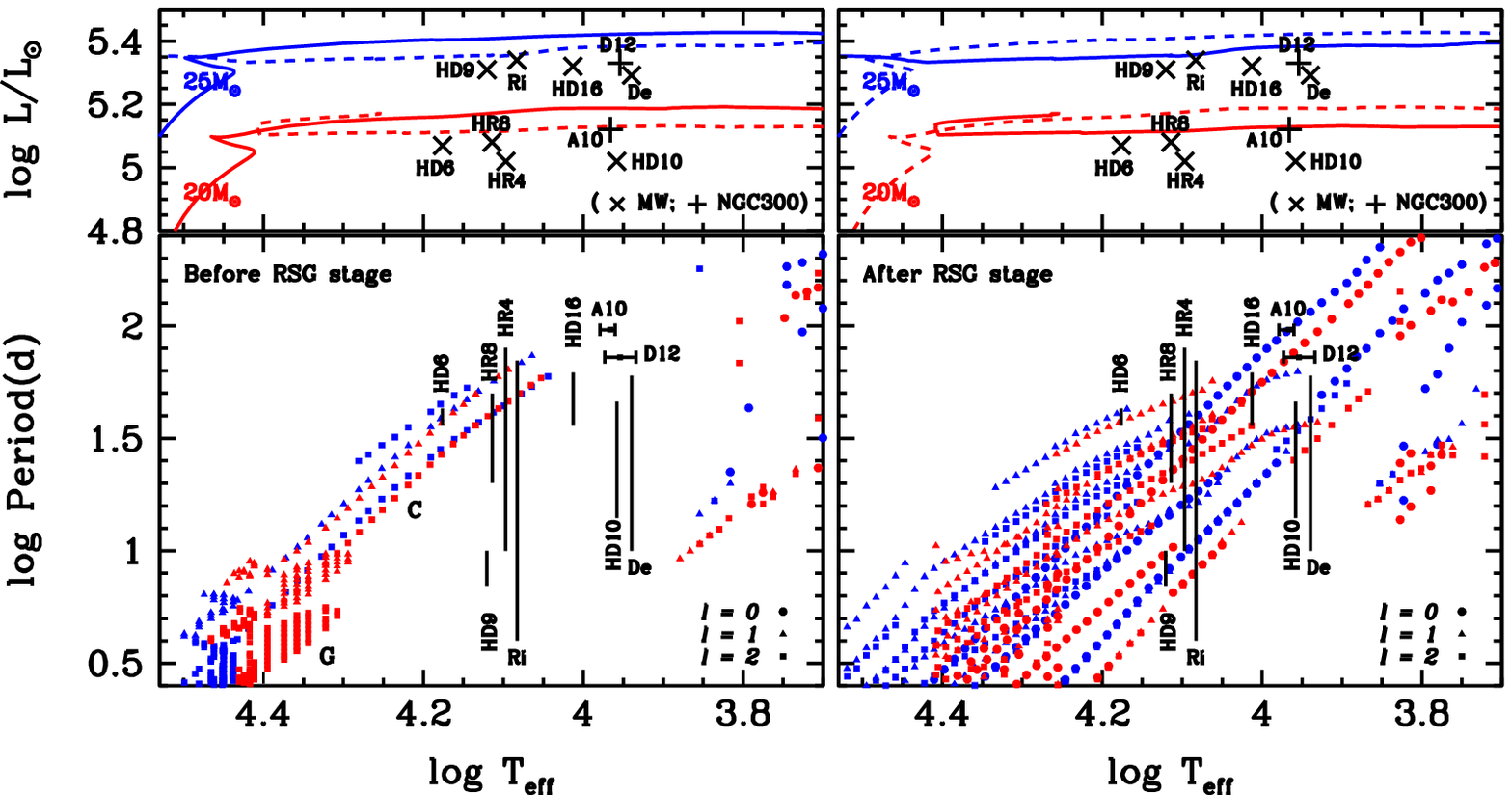}\hfill\includegraphics[height=.2\textheight]{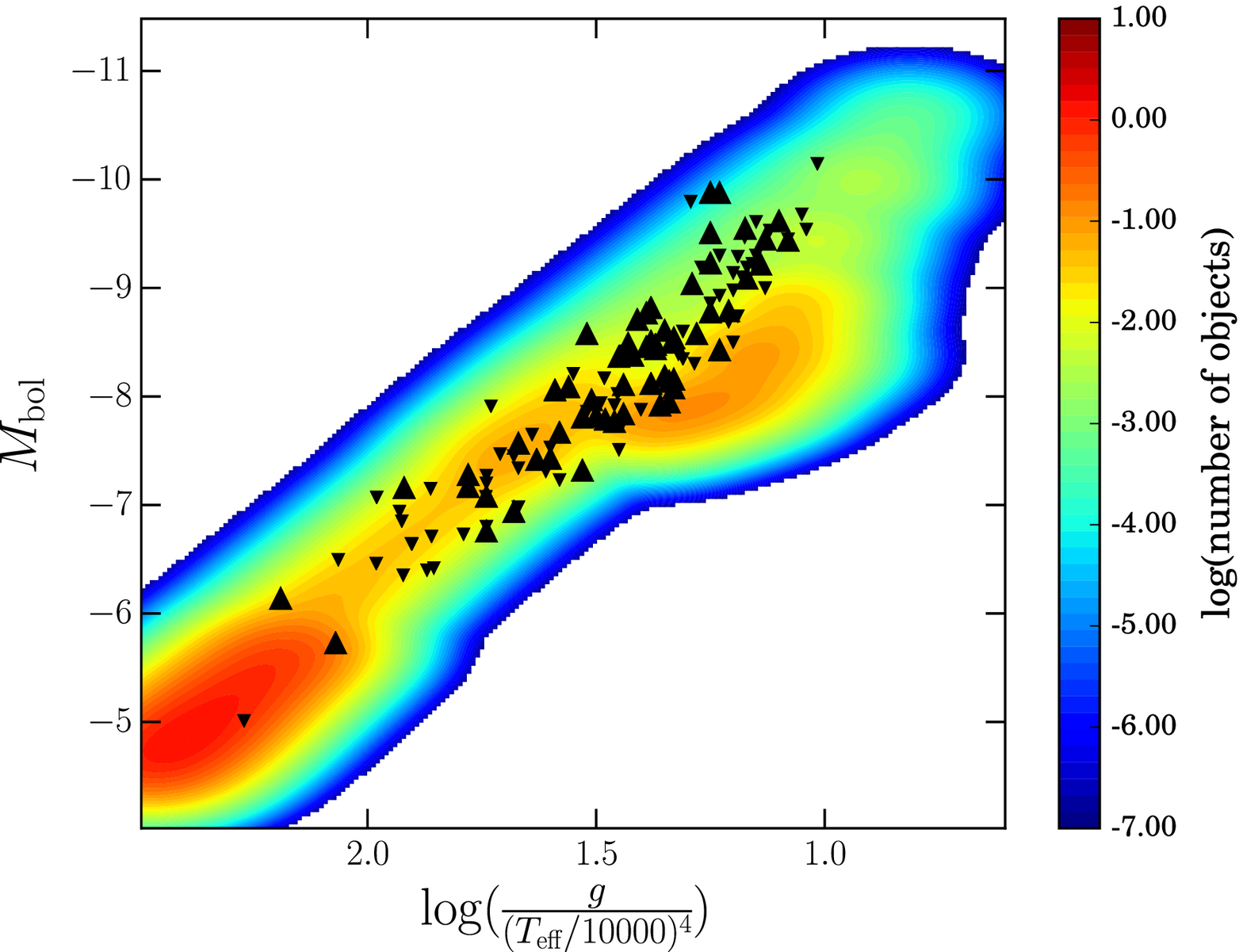}
\caption{\textit{Left and central panel:} HRD (on top) and excited pulsation periods visible at the surface of our models of the group 1 \textit{(left)} and group 2 \textit{(centre)}. Observed positions of some $\alpha$ Cyg variables are indicated. Adapted from \citet{Saio2013a}. \textit{Right panel:} Position of observed BSGs in the FWL relation plot, and expected positions of a synthetic population of rotating stars (coloured region). Adapted from \citet{Meynet2015b}.}
\label{Georgy_FigPulsa}
\end{center}
\end{figure}

A first consequence of this second crossing of the HRD is the co-existence of two populations of blue supergiants (BSGs), one from stars just after the main sequence, going to cross for the first time (group 1), and the other one from stars coming back from the red side of the HRD (group 2). The study of the pulsational properties of both groups shows that they are very different, and can thus be used to determine the membership \citep{Saio2013a}. Figure~\ref{Georgy_FigPulsa} shows the excited periods that are visible at the surface. Before the RSG phase (left panel), group 1 stars have only a few excited modes and almost none in the $\log(T_\text{eff})$ range from 4.0 to 4.2. On the other hand, group 2 stars (centre panel), that have lost a huge amount of mass during the RSG phase, have a much favourable $L/M$ ratio, that allows for much more modes to be excited. The predicted periods are in good agreement with the observed period of $\alpha$ Cygni variables. This group of BSGs can also be used to test the physics of convection in massive stars. We have shown in \citet{Georgy2014a} that using the Ledoux criterion instead of the Schwarzschild one provides surface abundances for $\alpha$ Cyg variables in better agreement with observations.

\citet{Kudritzki2008a} have shown that observed BSGs lie in a linear relation if they are plotted in the plane $\log(L)$ vs. $\log(g/T_\text{eff}^4)$ where $g$ is the surface gravity (see the stars in the right panel of Fig.~\ref{Georgy_FigPulsa}). This relation is called the flux weighted gravity-luminosity relation (FWLR). In \citet{Meynet2015b}, we have shown that our models are able to reproduce reasonably well this relation (right panel of Fig.~\ref{Georgy_FigPulsa}). However, we may generate too much group 2 stars, producing a slightly off-relation region where we expect from our models a high density of BSGs that are not observed. This could indicate that our models have an initial velocity that is too high compared to the average velocity of a typical massive star, or that our enhancement of the mass-loss rate during the RSG phase is too high.

\section{Conclusion}

The two examples discussed above show the consequences of the mass-loss rates during the RSG phase on the further evolution of the star. Observations of BSGs provide constraints on the mass-loss rate that have to be used in the stellar evolution models during that phase \citep[see also][]{Georgy2012a,Meynet2015a}. However, a better determination of the RSG mass-loss rates are extremely needed in order to improve the stellar modelling of the advanced stages of stellar life.

\vspace{.3cm}\noindent\textbf{Acknowledgments}

\noindent CG acknowledges support from EU-FP7-ERC-2012- St Grant 306901.

\bibliographystyle{astron} 
\bibliography{Georgy}

\vspace{.4cm}\noindent\textsc{Chiavassa:} Do you take into account the mass loss temporal variability during the RSG phase? As well as the chemical composition of the star?\\

\noindent\textsc{Georgy:} The mass-loss rate recipe for the RSG phase depends usually on stellar parameters, basically the temperature and luminosity. So the long-term variability due to the secular evolution of the star is accounted for through that way. The possible short term variability is not accounted for. Depending on the typical variation timescale, it could or could not be included in the stellar evolution codes: we need to have timesteps not too small. In the latter case, we would require a time-average mass-loss rates. Concerning the chemical composition, as far as I know, there is no mass-loss prescriptions that account explicitly for it in the literature.\\

\noindent\textsc{Moravveji:} We should not ignore the fact that to each datapoint on the FWLR contour plot, there should be attached to it a large error bar. So, in spite of that, is there a ``statistically'' significant difference between your predictions for rotating vs. non-rotating tracks?\\

\noindent\textsc{Georgy:} First of all, I would like to precise that we have included an error in our contour plot, that is 0.15 dex in magnitude, and 0.075 dex in $\log(T_\text{eff})$. We did not perform any statistical test to compare our predictions. They are slightly different, because the tracks are definitely different between rotating and non-rotating models. However, we are far to have a big enough observational sample to test the detail of the models so far.\\

\noindent\textsc{De Marco:} Ledoux criterion is \textit{very} different. It seems like we should have noticed that the Schwarzschild criterion was producing very different results. How likely is that this is the right explanation?\\

\noindent\textsc{Georgy:} At the beginning, we used the classical Schwarzschild criterion. But quickly, we had to face the problem that even if our prediction about the pulsation periods were relatively good, our predicted surface abundances were completely out \citep{Saio2013a}. So, we tried to change different things (overshoot, rotation, ...), but at the end, only the Ledoux criterion was able to reconcile both observations. However, and I want to be clear here, it does not mean that this is the correct solution. The treatment of convection in stellar codes is still very primitive, and it is well possible that the correct explanation is still different.\\

\noindent\textsc{Moffat:} Last week some of us were at the Wolf-Rayet workshop in Potsdam. At that meeting it was noted that from Geneva models, WR stars come from stars with $M_\text{ini}\gtrsim 25\,M_\odot$ while RSG come from $M_\text{ini}\lesssim 20\,M_\odot$, i.e. without overlap. Yet it was noted that the dense, rich cluster Westerlund 1 has a large population of both RSG and WR stars. What could be the reason to this?\\

\noindent\textsc{Georgy:} It is true that according to our models, the overlap is very short in duration (but still exists). One important point to remember here is that a non-negligible fraction of WR stars are probably produced in interacting systems. That could well be the case here.

\end{document}